\documentclass[10pt]{spie}  

\usepackage{amsmath,graphicx,bbm}
\usepackage{mathrsfs}
\usepackage{array}
\usepackage{subfigure}
\usepackage{ccaption}
\usepackage{color}
\usepackage{amssymb}
\usepackage{tikz}
\usepackage{float}
\usepackage{calc}
\usepackage{multirow}

\newcommand{\procspie}{Proc. SPIE } 
\newcommand{\pasp}{Publications of the Astronomical Society of the Pacific } 
\newcommand{\aap}{Astron. \& Astrophys. }

\newcommand{\cana}{\textsc{Canary} }
\newcommand{\sigdeux}[1]{\sigma^{2}_{\text{#1}}}

\newcommand{\para}[1]{\left(#1\right)}
\newcommand{\cro}[1]{\left[#1\right]}
\newcommand{\aver}[1]{\left\langle #1 \right\rangle}

\newcommand{\xth}[1]{#1^{\text{th}}}
\newcommand{\rz}{r_0}

\newcommand{\mx}{\text{max}\left( \text{PSF}_\text{sky}\right)}
\newcommand{\fcam}{F_\text{cam}}

\newcommand{\muncam}{\overline{n}_\text{cam}}
\newcommand{\se}{\sigma^2_\text{e}}
\newcommand{\boldr}{\boldsymbol{r}}
\newcommand{\boldrho}{\boldsymbol{\rho}}
\newcommand{\abeta}{\boldsymbol{a_\beta}  }
\newcommand{\aparabeta}{\boldsymbol{a_{\parallel\beta}} }
\newcommand{\hataparabeta}{\boldsymbol{\widehat{a}_{\parallel\beta}} }
\newcommand{\aperpbeta}{\boldsymbol{a_{\perp\beta}} }
\newcommand{\epsbeta}{\boldsymbol{\varepsilon_\beta}}
\newcommand{\epsparabeta}{\boldsymbol{\varepsilon_{\parallel\beta}} }
\newcommand{\hatepsparabeta}{\boldsymbol{\widehat{\varepsilon}_{\parallel\beta}} }
\newcommand{\sbeta}{\boldsymbol{s_\beta} }

\newcommand{\hatsbeta}{\boldsymbol{\widehat{s}_\beta} }
\newcommand{\mbeta}{\boldsymbol{m_\beta} }
\newcommand{\nbeta}{\boldsymbol{\eta_\beta}  }
\newcommand{\repsbeta}{\boldsymbol{r^\varepsilon_\beta}  }
\newcommand{\rbeta}{\boldsymbol{r_\beta}  }
\newcommand{\aalpha}{\boldsymbol{a_\alpha}  }

\newcommand{\salpha}{{\boldsymbol{s_\alpha}}  }
\newcommand{\aparaalpha}{\boldsymbol{a_{\parallel\alpha}}  }
\newcommand{\nalpha}{\boldsymbol{\eta_\alpha}  }
\newcommand{\ralpha}{\boldsymbol{r_\alpha}  }
\newcommand{\aepsparaalpha}{\boldsymbol{a^\varepsilon_{\parallel\alpha}}}
\newcommand{\eparabeta}{\boldsymbol{e_{\parallel\beta}}}
\newcommand{\aepsparabeta}{\boldsymbol{a^\varepsilon_{\parallel\beta}}}
\newcommand{\tmbeta}{\boldsymbol{\tilde{m}_\beta} }
\newcommand{\thataparabeta}{\boldsymbol{\tilde{\widehat{a}}_{\parallel\beta}} }
\newcommand{\tepsparabeta}{\boldsymbol{\tilde{\varepsilon}_{\parallel\beta}}}
\newcommand{\thatepsparabeta}{\boldsymbol{\tilde{\widehat{\varepsilon}}_{\parallel\beta}} }
\newcommand{\hol}{\tilde{h}_\text{ol}}
\newcommand{\hrej}{\tilde{h}_\text{rej}}
\newcommand{\tepsbeta}{\boldsymbol{\tilde{\varepsilon}_{\beta}} }
\newcommand{\taparabeta}{\boldsymbol{\tilde{a}_{\parallel\beta}} }
\newcommand{\taparaalpha}{\boldsymbol{\tilde{a}_{\parallel\alpha}} }
\newcommand{\tralpha}{\boldsymbol{\tilde{r}_{\alpha}} }
\newcommand{\trbeta}{\boldsymbol{\tilde{r}_{\beta}} }
\newcommand{\tnbeta}{\boldsymbol{\tilde{\eta}_{\beta}} }
\newcommand{\tnalpha}{\boldsymbol{\tilde{\eta}_{\alpha}} }
\newcommand{\taperpbeta}{\boldsymbol{\tilde{a}_{\perp\beta}} }
\newcommand{\thataparaalpha}{\boldsymbol{\tilde{a}_{\parallel\alpha}} }

\newcommand{\taepsparabeta}{\boldsymbol{\tilde{a}^\varepsilon_{\parallel\beta}}}

\newcommand{\teparabeta}{\boldsymbol{\tilde{e}_{\parallel\beta}}}
\newcommand{\trepsbeta}{\boldsymbol{\tilde{r}^\varepsilon_\beta}  }

\newcommand{\R}{\mathbf{R}}
\newcommand{\Zi}{\boldsymbol{Z}_i}
\newcommand{\Zj}{\boldsymbol{Z}_j}
\newcommand{\Mrz}{\mathbf{C}}
\newcommand{\Miz}{\mathbf{G}}

\newcolumntype{P}[1]{>{\centering\arraybackslash}p{#1}}

\title{PSF reconstruction validated using on-sky CANARY data in MOAO mode}

\author[a]{O.A. Martin}
\author[a]{C.M. Correia}
\author[b]{E. Gendron}
\author[b]{G. Rousset}
\author[b]{D. Gratadour}
\author[b]{F. Vidal}
\author[c]{T.J. Morris}
\author[c]{A.G. Basden}
\author[c]{R.M. Myers}
\author[a]{B. Neichel}
\author[a,d]{T. Fusco}

\affil[a]{Aix Marseille Universit\'e, CNRS, LAM, Laboratoire d'Astrophysique de Marseille, Marseille, France , 38 rue F. Joliot-Curie, 13388 Marseille Cedex 13, France}
\affil[b]{LESIA, Observatoire de Paris - Paris Sciences et Lettres, CNRS, Universit\'e Paris Diderot - Sorbonne Paris Cit\'e, Universit\'e P. et M. Curie - Sorbonne universit\'es, 5 pl. Janssen, 92190 Meudon,
France}
\affil[c]{Centre for Advanced Instrumentation, Durham Univ., South Road, Durham, DH1 3LE, UK}
\affil[d]{ONERA (Office National d'Etudes et de Recherches A\'erospatiales), B.P.72, F-92322 Ch\^atillon, France}

\authorinfo{Further author information: olivier.martin@lam.fr}

\begin{document} 
\maketitle
\begin{abstract}

In preparation of future Multi-Object Spectrographs (MOS) whose one of the major role is to provide an extensive statistical studies of high redshifted galaxies surveyed, the demonstrator \cana has been designed to tackle technical challenges related to open-loop Adaptive-Optics (AO) control with jointed Natural Guide Star~(NGS) and Laser Guide Star~(LGS) tomography.

We have developed a Point Spread Function~(PSF)-Reconstruction algorithm dedicated to MOAO systems using system telemetry to estimate the PSF potentially anywhere in the observed field, a prerequisite to post-process AO-corrected observations in Integral Field Spectroscopy~(IFS). 

In this paper we show how to handle off-axis data to estimate the PSF
using atmospheric tomography and compare it to a classical approach that
uses on-axis residual phase from a truth sensor observing a natural bright source.

We have reconstructed over 450 on-sky \cana PSFs and we get bias/1-$\sigma$ standard-deviation~(std) of 1.3/4.8 on the H-band Strehl ratio~(SR) with 92.3\% of correlation between reconstructed and sky SR. On the Full Width at Half Maximum~(FWHM), we get respectively 2.94 mas, 19.9 mas and 88.3\% for the bias, std and correlation. The reference method achieves 0.4/3.5/95\% on the SR and 2.71 mas/14.9 mas/92.5\% on the FWHM for the bias/std/correlation.

\end{abstract}


\section{Introduction}
\label{S:Intro}
For studies of formation and evolution of early galaxies, multiplexed observations of several objects need to be carried out at once using integral field spectrographs over a large Field-of-View (FoV). The need to correct only a limited number of discrete directions over the field led to the concept of Multi-Object adaptive optics (MOAO) which uses separate narrow field
wavefront correctors for each of several objects. To solve optical issues encountered on both wide FoV and large dimension systems, the MOAO concept consists in putting Wave-Front Sensors~(WFS) ahead of the Deformable Mirror~(DM) in the optical path. We adopt thus an \emph{open-loop configuration} since there is no feedback between DM and WFS.

Such a design has been proposed for \textsc{Eagle}~(\cite{Cuby2008}), the first multi-object Integral Field Units~(IFU) for the European-Extremely Large Telescope~(E-ELT)~(\cite{Pherson2012}), with a large FoV of 5 to 10 arc-minutes required for science observations~(\cite{Evans2008}). 

Today, \textsc{Mosaic} is considered the new MOAO-assisted multi-object IFU for the E-ELT~(\cite{Hammer2014}). The development of \textsc{Mosaic} will include a High-Multiplex Mode~(HMM), seeing-limited or coarsely-sampled using the Ground Layer Adaptive Optics~(GLAO) available on the E-ELT using the M4~(\cite{Vernet2012}) pre-focal adaptive mirror, for observing from 100 to 250 objects. In addition, it will also include a High-Definition Mode~(HDM) for observing from 10 to 20 faint galaxies with MOAO-corrected IFUs.

The feasibility of designing such a multi-channel instrument has to be demonstrated on-sky mainly for two reasons. The first relates to the reliability and high-fidelity control of DMs in open-loop configuration. Whereas the recursive nature of a closed-loop hides potential inaccuracies in the linear relationship between WFS and DM -- the interaction matrix -- in open-loop the absence of feedback puts to light ever-present deviations, DM creeping, variable gains, hysteresis to cite only a few. 
We build on the past experiments which have demonstrated operational open-loop DM control, e.g. \textsc{Volts}~\cite{Volt2008}, \textsc{Villages}~\cite{Gavel2008}, \textsc{Canary}~\cite{Gendron2011} and \textsc{Raven}~\cite{Lavigne2012}.

The second point relates to prove the reliability of the atmospheric open-loop tomography using both NGSs and LGSs, blending signals of different nature into a single estimation step. 

In order to progress on these topics \cana has been designed by an international consortium, involving the LESIA and the Durham University starting in 2007. The demonstrator was installed in 2010 at the William Herschel Telescope~(WHT) at La Palma, in the Canary islands. It has provided proof of our capability to handle the open-loop configuration and specify the on-sky performance, either using  NGSs only~(\cite{Vidal2014}), or mixed NGS/LGS~(\cite{Morris2014}). 

In Sect.~\ref{S:moao} we provide the analytic expression of the
residual phase in the science directions. With it we outline an error
breakdown of the main residual contributors, as tomography or
system bandwidth. 
We stress the fact that a successful attempt to reconstructing the PSF
is a direct consequence of a comprehensive understanding of the AO error budget.
 
In Sect.~\ref{S:PSFR}, we layout two methods to reconstruct the PSF:\\ 
\textit{i)} defining the analytic expression of the Optical Transfer Function~(OTF), computed from off-science guide-star telemetry. It relies on the tomographic reconstructor and on calibration matrices -- the same that are used during real-time operation -- leading us to dub the method \emph{RTC-based} PSF-R.\\

\textit{ii)} a reference method that uses \emph{Truth Sensor}~(TS) telemetry looking at bright science objects and thus directly measuring the residual phase. The implementation is very similar to the one proposed by V\'eran~(\cite{Veran1997}), with slight modifications required for MOAO systems. In normal operational conditions such sensor won't be usable since MOAO systems will generally observe very faint targets. 

Several approaches have been proposed to deal with the angular and focal anisoplanatism in LGS AO~(\cite{Flicker2003,Britton2006,Jolissaint2010}) based either on analytical anisoplanatism transfer functions~(\cite{Fusco2000}) or, in MCAO~(\cite{GillesJOSA2008}), using very accurate simulation tools. In the case of \cana both LGS and NGS are blended together in the tomography step -- a type we call integrated tomography -- using an MMSE reconstructor calibrated on-sky with the Learn~\&~Apply algorithm~(\cite{Vidal2011}). This yields a somewhat simpler formulation of the residuals in that we do not have to explicitly specify corrective terms for focal, angular or tilt anisoplanatism.
  
Results are shown using over 450 on-sky data sets, acquired in September 2013. We report in Sect.~\ref{S:results}, Strehl ratio~(SR) and Full Width at Half Maximum~(FWHM) evaluated on reconstructed PSF as function of sky quantities using focal plane data. We provide a statistical analysis to evaluate biases and variances using the RTC-based and TS-based method. We conclude in Sect.~\ref{S:conclusions}.

\section{Modelling phase residuals in MOAO systems}
\label{S:moao}

We now provide the analytic expressions for the tomographically
reconstructed on-axis residual phase
using off-axis measurements and later the same on-axis residual phase
from the truth sensor data.

\subsection{Estimating on-axis residual phase from off-axis measurements}

\begin{table}
	\caption{\textbf{Notation}}
	\scriptsize
	\begin{center}
		\begin{tabular}{c l}
			\hline
			\multicolumn{2}{c}{\textbf{Science directions} $\boldsymbol{\beta}$} \\ 
			\hline
			$\aparabeta$ & Atmospheric parallel phase on which the system is acting. Defined by the 36 first Zernike modes for \cana.\\
			& Appears in Eq.~\ref{E:eps_para_bi}.\\
			$\aperpbeta$ & Atmospheric high spatial frequencies not corrected. Appears in Eq.~\ref{E:eps_bi}.\\
			$\epsparabeta$ & Residual modes of the DM correction. Appears in Eqs.~\ref{E:eps_bi},~\ref{E:eps_para_bi}~\ref{E:sbeta},~\ref{E:varparabeta_hat} and~\ref{E:epspara}.\\
			$\mbeta$ & Modes produced by the DM. Appears in Eq.~\ref{E:v_bi} and~\ref{E:eps_para_bi}. \\
			$\sbeta$ & Concatenation of all WFS measurements in science directions, refer to the TS slopes for \textsc{Canary}.\\
			& Appears in Eqs.~\ref{E:sbeta},~\ref{E:varparabeta_hat} and~\ref{E:QSA}.\\ 
			$\nbeta$ & Additive noise contribution into WFS measurements~(TS for \textsc{Canary}). Appears in Eqs.~\ref{E:sbeta} and~\ref{E:varparabeta_hat}.\\
			$\rbeta$ & Additive aliasing contribution into WFS measurements~(TS for \textsc{Canary}). Appears in Eqs.~\ref{E:sbeta},~\ref{E:varparabeta_hat},~\ref{E:Dphiavg} and~\ref{E:rbeng}.\\
			$\repsbeta$ & Aliasing truly measured by on science directions WFS~(TS for \textsc{Canary}).\\& Defined in Eq.~\ref{E:rbeng} as the difference between $\rbeta$ and $\R\ralpha$. \\ 
			\hline
			\multicolumn{2}{c}{\textbf{Sensing directions} $\boldsymbol{\alpha}$} \\ 
			\hline
			$\salpha$ & Concatenation of all WFS measurements in sensing directions, here from off-axis WFS for the \cana case.\\& Appears in Eqs.~\ref{E:salpha} and~\ref{E:aparabeta}.\\ 
			$\aparaalpha$ & Measured modes by WFS in sensing directions. Appears in Eqs.~\ref{E:salpha} and~\ref{E:aparabeta}.\\
			$\nalpha$ & Noise contribution into WFS measurements. Appears in Eqs.~\ref{E:salpha},~\ref{E:aparabeta} and~\ref{E:epspara}. \\
			$\ralpha$ & Aliasing contribution into WFS measurements. Appears in Eqs.~\ref{E:salpha},~\ref{E:aparabeta},~\ref{E:rbeng},~\ref{E:epsrts} and~\ref{E:covepsrbe}.\\
			\hline
			\multicolumn{2}{c}{\textbf{Matrices}} \\ 
			\hline
			$\Mrz$ &  Zernike reconstruction matrix~(slopes to Zernike): calibrated on the bench.\\
			$\Miz$ &  Zernike interaction matrix~(Zernike to slopes): calibrated on the bench.\\
			$\R$ & Tomographic reconstructor~(slopes $\boldsymbol{\alpha}$ to slopes $\boldsymbol{\beta}$): calibrated on-sky using the Learn\& Apply approach.\\
			$\Zi(\boldsymbol{r})$ & $\xth{i}$ Zernike polynomials.\\
			\hline
			\multicolumn{2}{c}{\textbf{System}} \\ 
			\hline
			$g$ & Loop gain. \\ 
			$\Delta_t$ & Latency of the system in frames calibrated on the bench at 0.45 frame. \\ 
			$\nu_e$ & Sampling frequency of the RTC, 150 Hz\\
			
			\hline
			\multicolumn{2}{c}{\textbf{Miscellaneous}} \\ 
			\hline
			$\aver{\boldsymbol{x}}$ & Average of $\boldsymbol{x}$.\\
			$\Sigma_{\boldsymbol{x}\boldsymbol{y}}$ & Cross-covariance matrix of $\boldsymbol{x}$ and $\boldsymbol{y}$.\\
			$\boldsymbol{\widehat{x}}$ & Estimation of $\boldsymbol{x}$.\\
			$\boldsymbol{\tilde{x}}(z)$ & Z-transform of $\boldsymbol{x}(t)$.\\
			\hline  
		\end{tabular}
	\end{center}
	\label{T:notation}
\end{table}

\renewcommand{\arraystretch}{1.65}

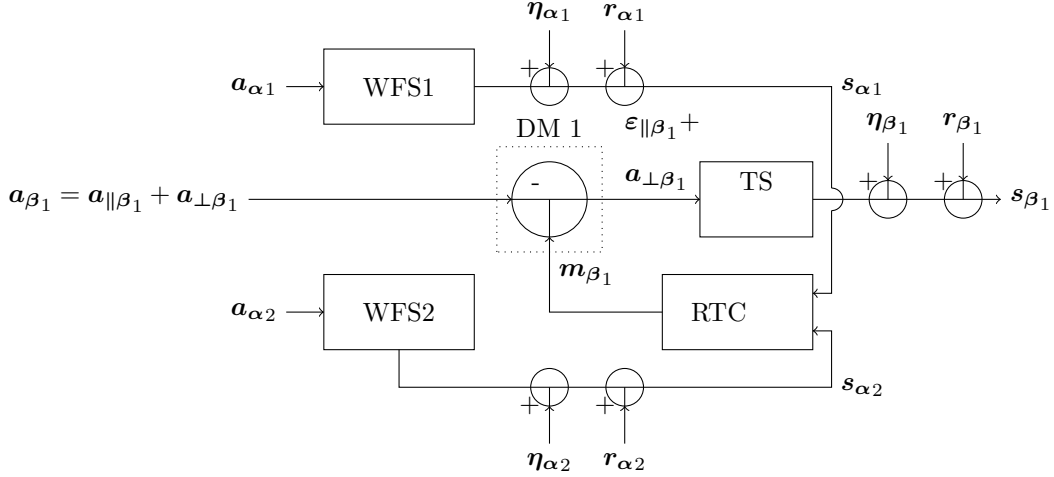
\begin{figure}[h!]
\begin{center}
	\begin{tikzpicture}{x=1cm,y=1cm,->=stealth}
	\draw (0,1) rectangle (2,2) ;\draw (1,1.5) node{WFS1};
	\draw (0,-1) rectangle (2,-2) ;\draw (1,-1.5) node{WFS2};
	\draw[->] (-.5,1.5) node[left]{$\aalpha_1$} -- (0,1.5);
	\draw[->] (-.5,-1.5) node[left]{$\aalpha_2$} -- (0,-1.5);
	\draw[->] (-1,0) node[left]{$\abeta_1 = \aparabeta_1 + \aperpbeta_1$} -- (2.5,0);
	\draw[->] (2.5,0) -- (5,0);
	\draw[->] (6.5,0) -- (9,0) node[right]{$\sbeta_1$};
	\draw (7.5,0) circle(0.25)node[above left]{+};\draw (8.5,0) circle(0.25)node[above left]{+};  
	\draw[->] (7.5,0.75) node[above]{$\nbeta_1$} -- (7.5,0.25); \draw (7.5,.25) -- (7.5,0);
	\draw[->] (8.5,0.75) node[above]{$\rbeta_1$} -- (8.5,0.25); \draw (8.5,.25) -- (8.5,0);
	
	\draw (3,-2.5) circle(0.25)node[below left]{+};\draw (4,-2.5) circle(0.25)node[below left]{+};
	
	\draw[->] (1,-2) -- (1,-2.5) -- (6.75,-2.5)  node[right]{$\salpha_2$} -- (6.75,-1.75) -- (6.5,-1.75);
	\draw (2,1.5) -- (6.75,1.5) node[right]{$\salpha_1$}  -- (6.75,.15);
	\draw (6.75,0.15) arc [radius=0.15, start angle=90, end angle= -90];
	\draw[->] (6.75,-.15) -- (6.75,-1.25) -- (6.5,-1.25);
	\draw[->] (3,-3.25) node[below]{$\nalpha_2$} -- (3,-2.75); \draw (3,-2.75) -- (3,-2.5);
	\draw[->] (3,2.25) node[above]{$\nalpha_1$} -- (3,1.75); \draw (3,1.75) -- (3,1.5);
	\draw[->] (4,-3.25) node[below]{$\ralpha_2$} -- (4,-2.75); \draw (4,-2.75) -- (4,-2.5);
	\draw[->] (4,2.25) node[above]{$\ralpha_1$} -- (4,1.75); \draw (4,1.75) -- (4,1.5);
	\draw (3,1.5) circle(0.25) node[above left]{+};\draw (4,1.5) circle(0.25)node[above left]{+};
	\draw (3,0) circle(0.5)node[above left]{-};
	\draw[fill=white] (5,-0.5) rectangle node[above]{TS} (6.5,0.5);
	\draw (4.5,-1) rectangle (6.5,-2) ;\draw (5.25,-1.5) node{RTC};
	\draw[->] (4.5,-1.5) -- (3,-1.5) -- (3,-.5);
	\draw (3,-.5) -- (3,0);
	\draw (4.5,0) node[above, text width=1cm]{$\epsparabeta_1 + \aperpbeta_1$};
	\draw (3,-1) node[right]{$\mbeta_1$};
	\draw[dotted] (2.3,-0.7) rectangle (3.7,0.7);
	\draw (3,0.7) node[above]{DM 1};
	\end{tikzpicture}
\end{center}
	\caption{\small{Description by block functions of a elementary MOAO system using two sensing WFS to compensate the turbulence into one science direction. See text for a more complete description.}}
	\label{F:moao}
\end{figure}

We illustrate in Fig.~\ref{F:moao} a schematic diagram of a single science channel MOAO system with two GS. Each of these sensing channels has a WFS, assumed to be a Shack-Hartman; it measures the uncompensated phase $\aalpha_j$ in the direction $\boldsymbol{\alpha}_j$. This WFS delivers noisy slopes assumed to be a linear function of the input phase through $\Miz$ -- the Zernike interaction matrix, calibrated on the bench, which gives the WFS slopes from the Zernike modal coefficients according to
\begin{equation} \label{E:salpha}
\salpha = \Miz\aparaalpha + \nalpha + \ralpha,
\end{equation}
with $\salpha$ the concatenation of WFS measurements, $\aparaalpha$ the Zernike modes corresponding to the parallel phase, $\nalpha$ and $\ralpha$ the additive measurement noise and aliasing respectively. The parallel phase here is defined as the Zernike modes the DM is able to correct for. From the calibration proposed by Vidal~(\cite{Vidal2014}), the parallel phase is composed up by the 36 first Zernike modes. On top of that, we assume the DM actuator pitch and the WFS sub-aperture pitch are identical and the fitting caused by modes higher than 36 is not sensed by the WFS -- unless in its aliased form.

The WFS slopes $\salpha$ are used to estimate the slopes $\hatsbeta$ in direction $\beta$ through the tomographic reconstructor $\R$ defined in slopes space~(\cite{Vidal2010}). The latter is calibrated on-sky. The estimation of the on-axis phase $\hataparabeta$ is then given by~:
\begin{equation} \label{E:aparabeta}
\begin{aligned}
\hataparabeta & = \Mrz\R\salpha\\
&\simeq \Mrz\R\Miz\aparaalpha + \Mrz\R\nalpha + \Mrz\R\ralpha
\end{aligned}
\end{equation}
where we use $\Mrz$ -- the pseudo-inverse of $\Miz$ -- to reconstruct the pupil-plane phase out of slopes. We further assume $\Mrz\Miz \simeq \mathbf{I}$, the identity matrix. The DM produces a corrective phase $\mbeta$, updated recursively from $\hataparabeta$. To reproduce the total system fractional delay $1+ \Delta_t$, with $\Delta_t$ calibrated on the bench to be $0.45$ of a frame, we combine values of $\hataparabeta$ acquired at time $t-1$ and $t-2$. Mathematically, we have~:
\begin{equation} \label{E:v_bi}
\mbeta(t) = (1-g)\times\mbeta(t-1) + 
 g \para{\Delta_t\hataparabeta(t-2) + (1-\Delta_t)\hataparabeta(t-1) }.
\end{equation}
In MOAO systems there is no "closed-loop" since there is no
WFS/DM feedback. To avoid confusion using the "closed-loop" and
"open-loop" terminology we adopted in the remainder of this paper
the term \emph{engaged loop} when $(g>0)$ meaning the loop is running and \emph{disengaged loop} when the DM is flat $(g=0)$.

Applying the Z-transform to Eq.~\ref{E:v_bi} gives
\begin{equation} \label{E:tmi}
\tmbeta(z) = \hol(z) \thataparabeta (z),
\end{equation}
where $z = e^{-2i\pi\nu/\nu_e}$, $\nu_e$ the temporal sampling frequency and $\hol(z)$ the controller transfer function defined as
\begin{equation} \label{E:hol}
\hol(z) = g\times \dfrac{\Delta_t + (1-\Delta_t)z}{z(z-1+g)}.
\end{equation}

We can now ascribe the phase residual as~:
\begin{equation} \label{E:eps_bi}
\epsbeta = \epsparabeta + \aperpbeta,
\end{equation}
where $\epsparabeta$ is the in-band residual leftover after DM compensation: 
\begin{equation} \label{E:eps_para_bi}
\epsparabeta = \aparabeta - \mbeta.
\end{equation}
whereas $\aperpbeta$ is the DM fitting error. 

Combining Eqs.~\ref{E:aparabeta},~\ref{E:tmi},~\ref{E:eps_bi} and \ref{E:eps_para_bi}, the Z-transform of $\epsbeta$, denoted $\tepsbeta(z)$, becomes:
\begin{equation} \label{E:tvbi1}
\tepsbeta = \taparabeta - \hol\Mrz\R\Miz\thataparaalpha -\hol\Mrz\R\tnalpha- \hol\Mrz\R\tralpha + \taperpbeta.
\end{equation}
where $\taparabeta - \hol\Mrz\R\Miz\thataparaalpha$ is the \emph{aniso-servo} error that combines both tomographic reconstruction and  system temporal lag errors. For on-sky performance diagnostic purposes, we have split this aniso-servo error into a sum of two components: a pure tomographic error plus a pure temporal error. We define the \emph{tomographic error} as follows:
\begin{equation} \label{E:ebe}
\teparabeta = \taparabeta - \Mrz\R\Miz\taparaalpha,
\end{equation}
which allows us to rewrite Eq.~\ref{E:tvbi1} as 
\begin{equation}\label{E:tvbi2}
\tepsbeta = \underbrace{\teparabeta}_\text{Tomography} + \underbrace{\hrej\Mrz\R\Miz\taparaalpha}_\text{Servo-Lag} - \underbrace{\hol\Mrz\R\tnalpha}_\text{Noise}
 - \underbrace{\hol\Mrz\R\tralpha}_\text{Aliasing} + \underbrace{\taperpbeta}_\text{Fitting},
\end{equation}
where the controller's rejection transfer function $\hrej$ is
\begin{equation} \label{E:hrej}
\hrej(z) =  1 - \hol(z).
\end{equation}

\begin{figure}[h!]
	\begin{center}
		\includegraphics[scale=.75]{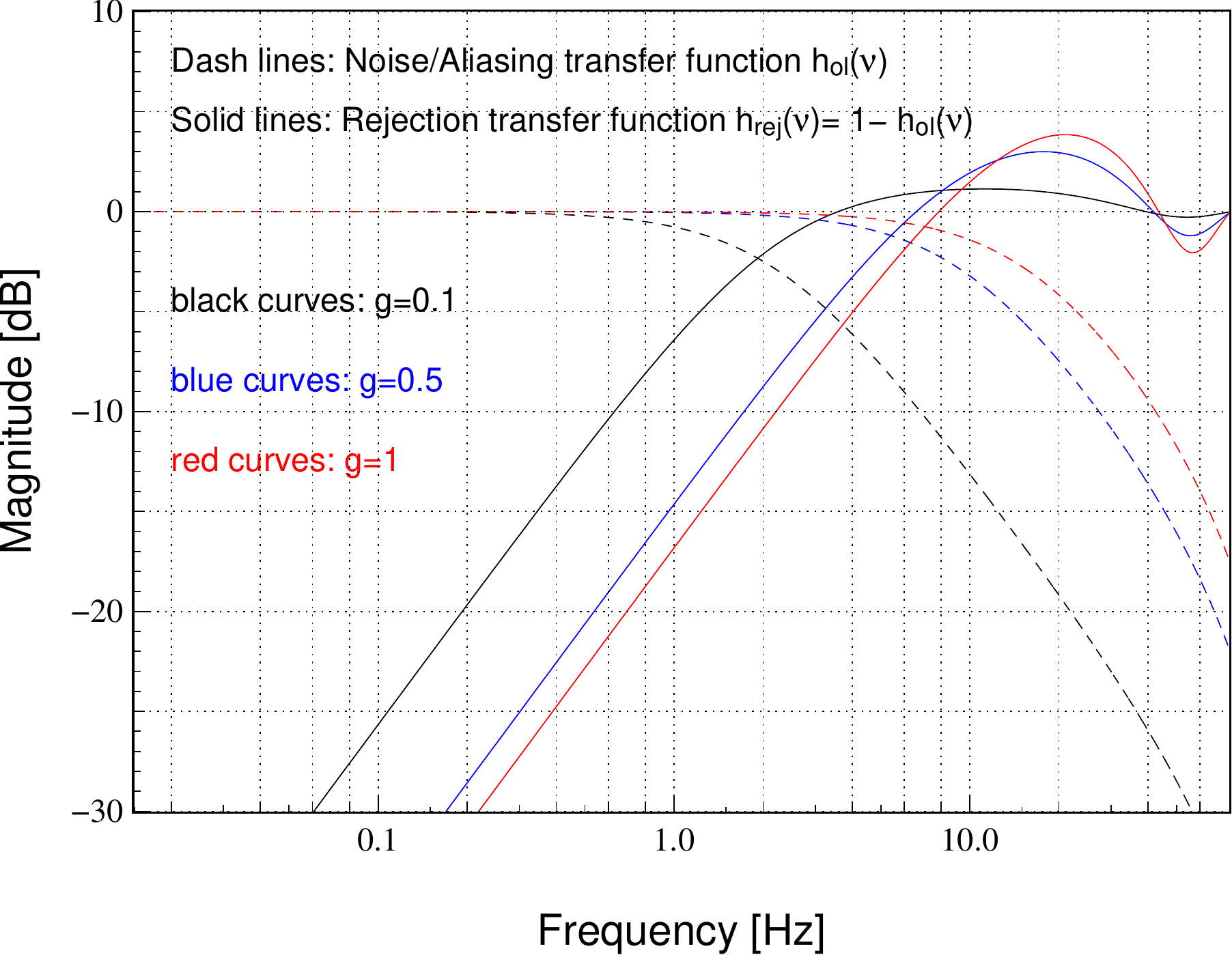}
	\end{center}
	\caption{\small{Bode magnitude plots of $\hrej$ and $\hol$ for three different values of gain, $\Delta_t = 0.45$ frames and  $\nu_e = 150$ Hz.}}
	\label{F:hrej}
\end{figure}

In Fig.~\ref{F:hrej}, we display Bode magnitude curves for both $\hrej$ and $\hol$ for three different gain values. According
to Eq.~\ref{E:tvbi2}, $\hol$ is the temporal transfer function operating on the off-axis aliasing and noise. We thus filter out the high temporal
frequencies of noise and aliasing, while the servo-lag error results essentially from the high temporal frequencies of the tomographic estimation since $\hrej$ does behave as a high-pass filter. We use the high-pass/low-pass features of the transfer functions to
compute the residual phase Wave-Front~(WF) error variance $\sigdeux{$\varepsilon$}$~:
\begin{equation} \label{E:var_eps}
\sigdeux{$\varepsilon$} \simeq \sigdeux{Tomography} + \sigdeux{Servo} + \sigdeux{Noise} 
+ \sigdeux{Alias} + \sigdeux{Fitting},
\end{equation}
where the WF error is given by summing over the variance of individual modes. To obtain Eq.~\ref{E:var_eps} from Eq.~\ref{E:tvbi2} the following approximations have been done~:
\begin{itemize}
\item All terms are considered statistically independent. More specifically the tomographic error have been separated from the servo-lag error. Mathematically, it means we neglect the correlation between $\teparabeta$ and $\hrej\Mrz\R\Miz\taparaalpha$. As shown in Fig.~\ref{F:hrej}, the signal filtered out by the transfer function $\hrej$ is depleted from its low temporal frequencies. On the other hand the tomographic error term follows a negative slope temporal spectrum: the bulk of the error is concentrated on the low temporal frequencies. These observations have led us to split both and disregard any cross term.
\item For the same reasons, we neglect the correlation between the aliasing $\hol\Mrz\R\tralpha$, mostly composed of low temporal frequencies, and the servo-lag error $\hrej\Mrz\R\Miz\taparaalpha$.
\item The correlation between the off-axis aliasing $\hol\Mrz\R\tralpha$ and the tomographic error $\teparabeta$ is neglected as well. The former is spatially filtered by the MMSE reconstructor and later temporally filtered by the loop transfer $\hol$. We thus expect the correlation of these two residual terms to be very weak prompting us to set it to null.
\item We further neglect the correlation between the high order modes $\taperpbeta$ above the DM cut-off frequency and the parallel ones included in the tomographic and servo-lag errors.
\item We assume the DM is well represented by Zernike modes, using the Zernike interaction matrix calibrated on the bench.
\end{itemize}

\subsection{On-axis residual phase measured by the Truth sensor}

\cana uses a \emph{Truth Sensor}, i.e. an extra WFS measuring the residual phase on-axis here $\boldsymbol{\beta}$. In practice, observing faint targets precludes the use of such sensor to probe the residual turbulence in the science directions of interest. As a technical demonstrator, \cana observes star asterisms composed of a bright central star, surrounded by up to two or three off-axis stars, usually separated by less than one arc-minute from the center~(\cite{Morris2014,Vidal2014}). The TS  is required for calibration and diagnostic purposes. Its vector of slopes $\sbeta$ -- see Fig.~\ref{F:moao} -- provides a measurement of the residual phase with additional noise and aliasing terms~:
\begin{equation} \label{E:sbeta}
\sbeta = \Miz\epsparabeta + \nbeta + \rbeta.
\end{equation} 
Using the Zernike reconstruction matrix $\Mrz$, we get~:
\begin{equation} \label{E:varparabeta_hat}
\begin{aligned}
\hatepsparabeta & = \Mrz\sbeta\\
& \simeq \epsparabeta + \Mrz\nbeta + \Mrz\rbeta.
\end{aligned}
\end{equation}
Pursuing as before we get~:
\begin{equation} \label{E:varparabeta_hat2}
\thatepsparabeta \simeq \underbrace{\teparabeta + \hrej\Mrz\R\Miz\taparaalpha - \hol\Mrz\R\tnalpha}_{\tepsparabeta} +  \Mrz\tnbeta 
 + \Mrz\underbrace{(\trbeta -\hol\R\tralpha)}_{\trepsbeta}. 
\end{equation}
Equation~\ref{E:varparabeta_hat2} shows that the TS measures a residual aliasing term, i.e. the difference between the pure TS aliasing $\trbeta$ and the post-tomographic term $\hol\R\tralpha$. We define $	\repsbeta$ as this \emph{aliasing residual} measured by the TS during observation as~:
\begin{equation} \label{E:rbeng}
	\repsbeta \simeq \rbeta - \R\ralpha.
\end{equation}
In Eq.~\ref{E:rbeng}, we do not consider the temporal filtering of the off-axis aliasing as it appears in Eq.~\ref{E:varparabeta_hat2}, which leads to an over-estimation of the wavefront error, thus an under-estimation of the SR.

\section{PSF reconstruction in MOAO}\label{S:PSFR}					

In tomographic systems the challenge is to be able to estimate the AO residuals without actual measurements of the residual phase in the directions of interest.  For convenience we use Optical Transfer Functions~(OTF). PSFs are readily computed from OTFs by Fourier transforms.

In the following, we propose to compare two different techniques:

\begin{itemize}
	\item \textbf{RTC-based method:} this is the adaptation of the
          V\'eran's method~(\cite{Veran1997}) to MOAO systems
          employing atmospheric tomography.  It
          consists in estimating the on-axis residual phase that would
          be otherwise measured by the TS when the loop is
          engaged. To than end, we generalize calculations presented
          in Sect.~\ref{S:moao} to estimate the covariance matrix of
          the residual in-band AO modes.
	\vspace{.25cm}
	\item \textbf{TS-based method:} Here we use a slightly
          modified classical PSF-R algorithm to account for specifics
          of propagated error terms using the Truth sensor
          telemetry. On the \cana demonstrator, the TS is used on-axis
          with a bright star for demonstrating operational MOAO plus
          diagnosing performance. It is used here as a reference
          method but will not be available in normal MOAO operations.
\end{itemize}

\subsection{OTF decomposition }

The OTF of an AO system is commonly modelled as a multiplication of the telescope OTF including static aberrations, and an AO OTF of the residuals~(\cite{Veran1997}). The latter is then split into an in-band \textit{parallel part}, OTF$_{\epsparabeta}$ over spatial frequencies the AO systems has acted upon and an \textit{orthogonal part}, OTF$_{\perp\beta}$ that gathers all the spatial frequencies above the DM cut-off frequency. The final  OTF$_{\epsbeta}$, can be written as follows:
\begin{equation} \label{E:otf_res}
\text{OTF}_{\epsbeta} = \text{OTF}_{\text{NCPA}_\beta} \times \text{OTF}_{\text{Static}_\beta} \times \text{OTF}_{{\perp\beta}} \times \text{OTF}_{\epsparabeta},
\end{equation}
where $\text{OTF}_{\text{NCPA}_\beta}$ is the Non Common Path
Aberrations~(NCPA) including the telescope OTF in directions
$\boldsymbol{\beta}$. Since we include the telescope OTF with the NCPA
term the remaining terms must be free from it to avoid incorporating it
more than once.

\subsection{Determination of in-band residual phase: OTF$_{\epsparabeta}$ }

The long-exposure OTF is given by \cite{Veran1997}:
\begin{equation} \label{E:OTF_para}
\text{OTF}_{\epsparabeta}(\boldrho/\lambda) = e^{-0.5\times \bar{D}_{\phi_{\epsparabeta}}(\boldrho)},
\end{equation}
where $D_{\phi_{\epsparabeta}}$ is the residual phase Structure Function~(SF) which when averaged over the telescope pupil becomes~:
\begin{equation}
\bar{D}_{\phi_{\epsparabeta}}(\boldrho) = \dfrac{\iint \aver{\para{\phi_{\epsparabeta}(\boldr) - \phi_{\epsparabeta}(\boldrho + \boldr)}^2}P(\boldr)P(\boldrho+\boldr)d\boldr}{\iint P(\boldr)P(\boldrho+\boldr)d\boldr},
\end{equation}
where $P(\boldr)$ the pupil function.  Using the 36 first Zernike polynomials we get~:
\begin{equation}
\phi_{\epsparabeta}(\boldr) = \sum_{i=2}^{36} \epsparabeta\Zi(\boldr).
\end{equation}
The residual phase structure function then becomes~:
\begin{equation} \label{E:Dphiavg}
\bar{D}_{\phi_{\epsparabeta}}(\boldrho) = \sum \limits_{i,j}^{n} \aver{\epsparabeta(i) \epsparabeta(j)} U_{ij}(\boldrho),
\end{equation}
where~:
\begin{equation} \label{E:Uij}
U_{ij}(\boldrho) = \dfrac{\iint P(\boldr)P(\boldrho + \boldr)d\boldr\para{\Zi(\boldr) - \Zi(\boldr + \boldrho)} \times 
\para{\Zj(\boldr) - \Zj(\boldrho + \boldr)} d^2\boldr} {\iint P(\boldr)P(\boldrho + \boldr)d^2\boldr}
\end{equation}
From Eq.~\ref{E:Dphiavg}, the estimation of the PSF boils down to estimating the covariance matrix of the residual phase expressed on the Zernike modal basis $\Sigma_{\epsparabeta \epsparabeta}(i,j) =  \aver{\epsparabeta(i) \epsparabeta^t(j)}$.

\subsubsection{RTC-based method}
\label{SSS:rtc}

The RTC-based method consists in determining $\epsparabeta$ from the
off-axis WFS telemetry. Eq.~\ref{E:tvbi2} gives an analytic expression
of the Z-transform of the residual modes $\epsbeta$. In subtracting
off the orthogonal part $\aperpbeta$ (i.e. the fitting) and in transforming back to the real space, we get:
\begin{equation} \label{E:epspara}
\epsparabeta  \simeq \eparabeta + \aepsparabeta + \Mrz\R(h_\text{ol}*\nalpha) +\Mrz\R\ralpha,
\end{equation}
where $h_\text{ol}$ the impulse response of the MOAO system, $(.*.)$ the convolution product and $\aepsparaalpha$ is the output of a first order low-pass filter defined by: 
\begin{equation}
\taepsparabeta(z) = \hrej(z) \times \Mrz\R\Miz\boldsymbol{\widehat{\tilde{a}}_{\parallel\alpha}}(z),
\end{equation}
where $\hrej$ is given by Eq.~\ref{E:hol} and Eq.~\ref{E:hrej}, and $\widehat{\tilde{a}}_{\parallel\alpha}$ is our best estimates of the off-axis parallel noise-free modes. From Eq.~\ref{E:epspara} and Eq.~\ref{E:ebe}, assuming all the terms
involved in Eq.~\ref{E:epspara} are independent processes, we get the
following expression for the spatial covariance of the residual phase
in the direction $\beta$~:
\begin{equation} \label{E:sigepsbertc}
\Sigma_{\epsparabeta\epsparabeta } = \Sigma_{\eparabeta\eparabeta}+ \Sigma_{\aepsparabeta\aepsparabeta} + p_\eta\times\Mrz\R\Sigma_{\nalpha\nalpha}\R^t\Mrz^t
 +\Mrz\R\Sigma_{\ralpha\ralpha}\R^t\Mrz^t,
\end{equation}
where~:
\begin{itemize}
    \item $\Sigma_{\eparabeta\eparabeta} = \aver{\eparabeta\eparabeta^t}$ is the tomographic error covariance matrix. Without TS, the tomographic error as defined in Eq.~\ref{E:ebe} can not be evaluated empirically from the telemetry. Nevertheless, a model of this covariance matrix is described in \cite{Martin2016L3S} based on the tomographic reconstructor $\R$ and relevant parameters which are~:
    \begin{itemize}
        \item $n_l$: Number of considered discrete layers, set to five for \cana
        \item $h_l$: altitude of turbulent layers
        \item $r_0(h_l)$ profile: The $r_0$ value and the fractional $r_0$ per layers
        \item $L_0(h_l)$: $L_0$ value for each layer
        \item NGS/LGS WFS angular positions: they are sufficiently well known using encoders information from the target acquisition system~(\cite{Gendron2011,Morris2014})
        \item LGS focus distance and depth: calibrated on-sky to respectively 21~km and 1.5~km~(regardless of the airmass)~(\cite{Morris2014})
	    \item Off-axis WFS/TS mis-registration~(pupil shifts, rotation and magnification): calibrated on the bench
	    \item Telescope tracking error and vibrations: modeled by an isoplanatic tip-tilt over all NGS WFS
        \item Noise slope variance for each WFS, separately identified on the $\salpha$ time series by using a parabolic fitting on the temporal auto-correlation function.
    \end{itemize}
    Using the Learn~\&~Apply approach~(\cite{Vidal2011}), we retrieve this list of parameters in fitting our model over the covariance matrix of measured slopes. We then extrapolate the TS/off-axis WFS covariance matrix of slopes to retrieve the covariance matrix of the tomographic error $\Sigma_{\eparabeta\eparabeta}$~(\cite{Gendron2014}).
    \item $\Sigma_{\aepsparaalpha\aepsparaalpha} = \aver{\aepsparaalpha\aepsparaalpha^t} $ is the covariance matrix of the temporally filtered atmospheric parallel modes by the MOAO system. It is readily derived from the time-series of the off-axis WFS measurements and parameters from the loop during the observation.
    \item $\Sigma_{\nalpha\nalpha}$ is the noise covariance matrix of off-axis WFS telemetry, identified by a parabolic interpolation, near to a zero-delay, of temporal auto-covariance of $\salpha$, and $p_\eta = \dfrac{g}{2-g}(1-2g\Delta_t(1-\Delta_t))$ the noise propagation factor computed by~\cite{Vidal2014}.
    \item $\Sigma_{\ralpha\ralpha}$ is the covariance matrix of the off-axis aliasing derived from the model described in~\cite{Martin2016L3S} depending on the same list of parameters as described before for the computation of $\Sigma_{\eparabeta\eparabeta}$. 
\end{itemize}

 Contrary to approaches proposed by~(\cite{Britton2006,GillesJOSA2008}), using separate tilt measurements, we do not compute specific OTFs for either tip-tilt or high order modes. This is the role of the integrated tomography MMSE reconstructor blending measurements from both NGS and LGS to estimate the on-axis phase in the minimal mean square error sense~(\cite{Martin2016A&A}). 

\subsubsection{TS-based method}
\label{SSS:ts}

The SF in Eq.~\ref{E:Dphiavg} requires the covariance matrix of parallel modes $\Sigma_{\epsparabeta\epsparabeta}$ estimated from $\hatepsparabeta(\boldrho,t)$. Using Eq.~\ref{E:varparabeta_hat}, we get~:
\begin{equation} \label{E:sigvareps_hat}
    \Sigma_{\hatepsparabeta\hatepsparabeta} \simeq  \Sigma_{\epsparabeta\epsparabeta} +  \Mrz\Sigma_{\nbeta \nbeta}\Mrz^t + \Mrz\Sigma_{\repsbeta\repsbeta}\Mrz + \Sigma_{\epsparabeta\repsbeta}\Mrz^t + \Mrz\Sigma_{\repsbeta\epsparabeta}.
\end{equation}
From Eq.~\ref{E:rbeng}, the covariance matrix $\Sigma_{\repsbeta\repsbeta} = \aver{\repsbeta\repsbeta^t}$ can be derived as follows:
\begin{equation} \label{E:covrbe}
\Sigma_{\repsbeta\repsbeta} = \Sigma_{\rbeta\rbeta} + \R\Sigma_{\ralpha\ralpha}\R^t -\Sigma_{\rbeta\ralpha}\R^t -\R\Sigma_{\ralpha\rbeta},
\end{equation}
where the matrix $\Sigma_{\rbeta\ralpha}$ is the concatenation of the covariance matrices of off/on axis aliasing in loop disengaged~$(g=0)$. Assuming the independence of terms given in Eq.~\ref{E:tvbi2}, the TS aliasing $\repsbeta$ is only correlated with $\epsparabeta$. We thus have:
\begin{equation}\label{E:epsrts}
\Sigma_{\epsparabeta \repsbeta} \simeq \Mrz\aver{\R \ralpha \repsbeta^t}\Mrz^t.
\end{equation}
Combining Eq.~\ref{E:epsrts} with Eq.~\ref{E:rbeng} leads to:
\begin{equation} \label{E:covepsrbe}
\begin{aligned}
\Sigma_{\epsparabeta \repsbeta} &\simeq \Mrz\aver{\R \ralpha (\rbeta - \R\ralpha)^t}\Mrz^t\\
&\simeq \Mrz\para{\R \Sigma^t_{\ralpha\rbeta} - \R\Sigma_{\ralpha\ralpha}\R^t}\Mrz^t.
\end{aligned} 
\end{equation}

Finally, in combining Eq.~\ref{E:sigvareps_hat}, Eq.~\ref{E:covrbe} and Eq.~\ref{E:covepsrbe}, one gets the final expression of $\Sigma_{\boldsymbol{\varepsilon_{\parallel\beta}}\boldsymbol{\varepsilon_{\parallel\beta}}}$:
\begin{equation} \label{E:covepseps_ts}
\Sigma_{\epsparabeta \epsparabeta } \simeq \Sigma_{\hatepsparabeta\hatepsparabeta} - \Mrz\Sigma_{\nbeta \nbeta}\Mrz^t - \Mrz\Sigma_{\rbeta\rbeta}\Mrz^t + \Mrz\R\Sigma_{\ralpha\ralpha}\R^t\Mrz^t,
\end{equation}
where:
\begin{itemize}
\item $\Sigma_{\hatepsparabeta\hatepsparabeta} = \aver{\hatepsparabeta\hatepsparabeta^t}$ is the covariance matrix estimated from the times series of $\sbeta(\boldrho,t)$~(TS slopes) acquired on-sky. 
\item $\Sigma_{\nbeta \nbeta}$ is the noise covariance matrix estimated from the temporal auto-correlation function of $\sbeta$, using the same approach developed for $\Sigma_{\nalpha \nalpha}$ .
\item $\Sigma_{\rbeta\rbeta}$ is the covariance matrix of the TS aliasing in loop disengaged. As for $\Sigma_{\ralpha\ralpha}$, we do not need to consider any time-filtering of the aliasing, meaning we readily estimate this covariance matrix using the model proposed in~\cite{Martin2016L3S} depending on the list of parameters described before.
\item $\Sigma_{\ralpha\ralpha}$ is the same off-axis aliasing covariance matrix required for the RTC-based method.
\end{itemize}

Note the TS-based method does not include angular/focal anisoplanatism since the science object is also the reference star observed by the TS. In passing, note that unlike in V\'eran's approach~(\cite{Veran1997}) for closed-loop systems, the TS is here only an observer of the residual phase. The TS slopes are never used to update the DM commands. This is a noticeable difference in how the aliasing is handled: the TS aliasing is no longer anti-correlated to the low order modes produced by the DM to null out the WFS measurements, since there is no TS/DM feedback. Using the large bandwidth approximation, we would have $\Sigma_{\epsparabeta \repsbeta} = -\Sigma_{\repsbeta \repsbeta}$ in Eq.~\ref{E:covepsrbe}, which is not the case here. 

\subsection{NCPA OTF: $\text{OTF}_{\text{NCPA}_\beta}$} \label{SS:otfncpa}

The NCPA are calibrated on the bench, without turbulence and in closed loop on the TS, using the phase diversity algorithm proposed by Gratadour~(\cite{Gratadour2013b}). 
The residual phase is retrieved from a set of in/out focus images in the focal plane. The calibration leads to non-compensated high order modes and a fitting residual on the low order modes, reaching 105~nm rms of wavefront error. We consider these NCPA as static during each observation runs~(three nights to six consecutive nights).

\subsection{Static OTF: $\text{OTF}_{\text{Static}_\beta}$} \label{SS:otfstat}

In addition to NCPA, we have to deal with common static aberrations $\aver{\abeta}$ in the science directions. These are not strongly mitigated by the loop as it is done for closed-loop systems. In MOAO, they must be calibrated on-sky and compensated by a DM offset. 

Unfortunately, these static aberrations are drifting slowly in
time because the field rotation and
the DM creeping effect ~(\cite{Kellerer2012}) and range 100~nm rms of wavefront error~(\cite{Martin2016A&A}). They are calibrated in loop-engaged using the on-going test tomographic reconstructor by averaging the TS telemetry with time~(see Eq.~\ref{E:QSA}) and applying a DM offset using the command matrix. This procedure is repeated several times over the night to tackle the DM creeping in the best way possible.

Even if the science objects are too faint for WF sensing, the \cana experiment has demonstrated it is feasible to offset the telescope
towards a bright star~(R $<$ 12) to get the required calibration~(\cite{Gratadour2013c}). For MOAO PSF-R purposes we determine $\aver{\abeta}$ from the TS measurements using the Zernike reconstruction matrix $\Mrz$, considering noise and aliasing as zero-mean processes, as follows:
\begin{equation} \label{E:QSA}
\aver{\abeta} = \Mrz\aver{\sbeta}.
\end{equation}

Equation~\ref{E:otf_res} involves a static OTF$_{\text{Static}_\beta}$ term. To avoid including multiple times the telescope OTF which is factored in every term as well as in  Eq.~\ref{E:QSA}, precaution must be taken by dividing the OTFs by that of the telescope \cite{Gendron2006} as follows:
\begin{equation} \label{E:OTFstat}
\text{OTF}_{\text{Static}_\beta}(\boldrho/\lambda) = \dfrac{ \iint_\mathcal{P} P(\boldr) P(\boldr + \boldrho) \times  e^{-i\aver{\abeta}(\boldr)} \times e^{-i\aver{\abeta}(\boldrho + \boldr)} d\boldr }{\text{OTF}_\text{tel}(\boldrho/\lambda)},
\end{equation}
where:
\begin{equation}
\text{OTF}_\text{tel}(\boldrho/\lambda) = \iint_\mathcal{P} P(\boldr) P(\boldr + \boldrho) d\boldr
\end{equation}
and applying a threshold for values of $\boldrho$ nearing the
telescope pupil to avoid division by zero for larger spatial frequencies than $D/\lambda$. 

\subsection{Fitting OTF: OTF$_{\perp\beta}$}

The DM fitting $\text{OTF}_{\perp\beta}$ is computed analytically from the Von-K\'arm\'an Power Spectrum Density~(PSD) of the atmosphere, which scales with $(d/\rz)^{5/3}$, set to zero in the band of frequencies lower than $1/2d$. The proportionality has been calibrated using end-to-end simulations~(\cite{Vidal2014}). We get $\sigdeux{Fitting} = 0.3125\times (d/\rz)^{5/3}$ for the \cana DM.

\section{PSF-R results using on-sky CANARY data}
\label{S:results}

\subsection{Statistics on reconstruction performance}
\label{SS:statistics}

\begin{figure}[H]
	\begin{center}
	\vspace{-3cm}
	
\hspace{-2.4cm} \includegraphics[scale=.9]{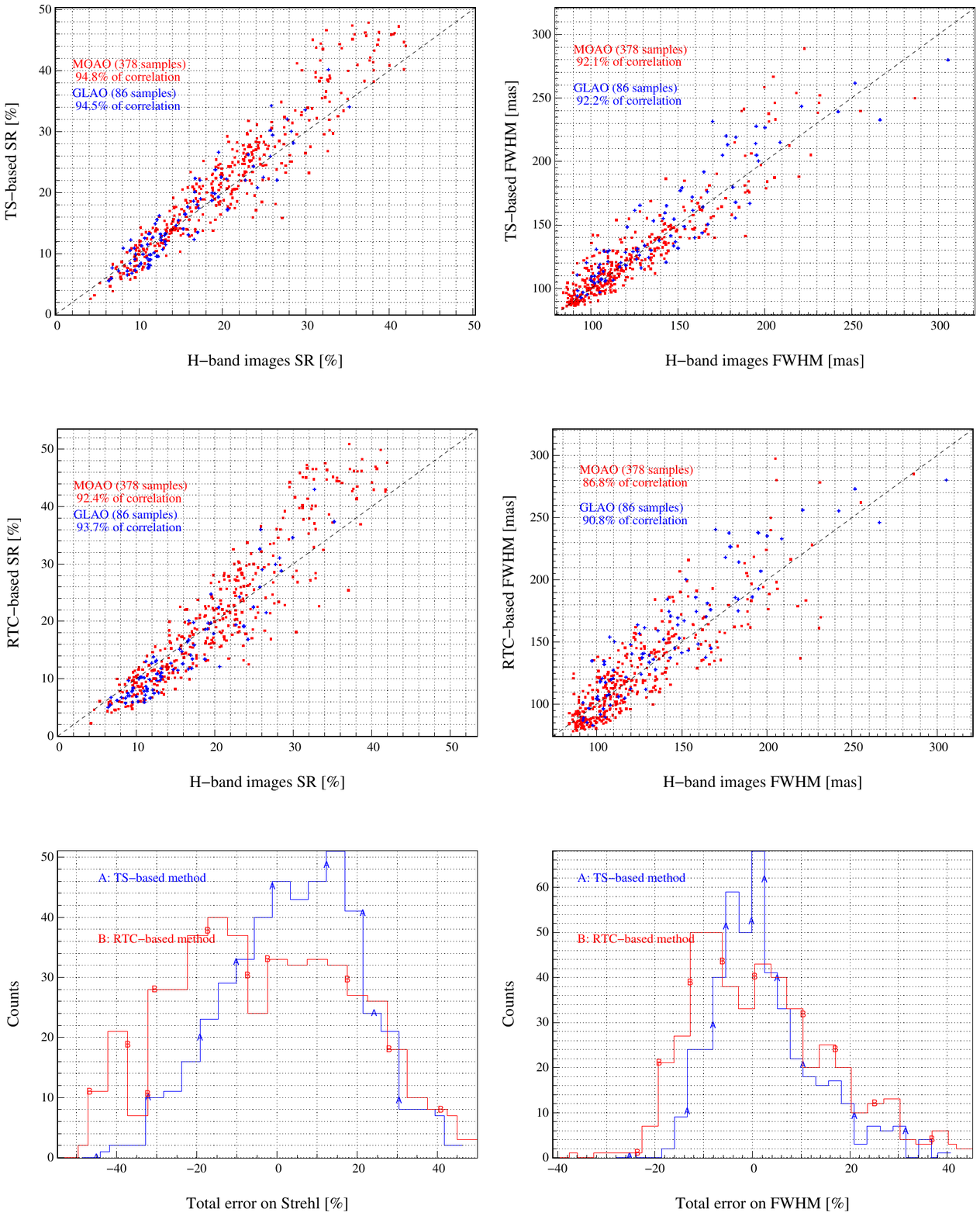}
\vspace{-5cm}

	\end{center}
	\caption{\small{\textbf{Left:} \textbf{a}) TS-based reconstructed SR v H-band sky SR, \textbf{b)} RTC-based reconstructed SR and \textbf{c}) histograms of the signed relative error on the SR. \textbf{Right:}  \textbf{a}) TS-based reconstructed FWHM as function of H-band sky FWHM, \textbf{b}) RTC-based reconstructed FWHM v sky one and \textbf{c)} histograms of the signed relative error on the FWHM. }}
	\label{F:stat}
\end{figure}

Figure~\ref{F:stat} depicts results on reconstructed Strehl-ratio~(SR) and Full Width at Half Maximum~(FWHM) as function of the corresponding sky values estimated on focal plane images. Each point is a result of the PSF-R performed on a single data set acquired in September 13, 15, 17 and 18 2013, where \cana was operating is phase B configuration, observing a four stars asterism, with a central bright star and with four Rayleigh LGS focused at 21~km~(see.~\cite{Morris2014} for observation details). We report in Table~\ref{T:stat} statistics of the reconstruction in terms of SR and FWHM.

\begin{table}
	\begin{center}
		\begin{tabular}{|c||c|c|c|c|}
			\hline
			Method & Criteria & Mean & Std & Correlation [\%]\\
			\hline
			\multirow{2}{2cm}{RTC-based}  & $\Delta$SR = SR$_\text{rtc}$ - SR$_\text{sky}$    & 1.29 & 4.82 & 92.3 \\
			& $\Delta$FWHM = FWHM$_\text{rtc}$ - FWHM$_\text{sky}$ [mas] &  2.94  & 19.9 & 88.3 \\
			
			\hline
			\multirow{2}{2cm}{TS-based}   & $\Delta$SR = SR$_\text{ts}$ - SR$_\text{sky}$   & 0.41 & 3.52  & 95.0 \\
			& $\Delta$FWHM = FWHM$_\text{ts}$ - FWHM$_\text{sky}$ [mas] & 2.71  & 14.9 & 92.5\\
			\hline		
		\end{tabular}
	\end{center}
	\caption{\small{Statistics on PSF-R performed using either the RTC-based or the TS-based method over 450 \cana MOAO data sets acquired in September 2013. SR range between 0 and 100.}}
	\label{T:stat}
\end{table}

H-band sky images were generally 1~s exposure time and averaged over 15 realizations. After subtracting the background calibrates on-sky, all spatial frequencies in the OTF higher than $D/\lambda$  are set to zero. We then evaluate $\muncam$ and $\se$, respectively the median and variance of the residual background noise, from the outer region part of inscribed circle of the image, then the flux $\fcam$ from the inside part. The SR is then estimated using the method proposed by Gendron~\cite{Gendron2011}:
\begin{equation} \label{E:SRest}
    \widehat{\text{SR}}_\text{sky} = \dfrac{\mx-\muncam}{a\times \fcam},
\end{equation}
where $a = \pi^2/4 (D^2-o^2)\times p^2/\lambda^2$, with $D$ and $o$
respectively the telescope and central obscuration diameter, and $p$ the camera pixel scale. According to Eq.~\ref{E:SRest}, the standard-deviation of the SR estimates can be derived as follows:
\begin{equation}
\begin{aligned}
    \sigdeux{\text{SR}} = & \para{\dfrac{\mx}{a\times \fcam}}^2\cro{ \dfrac{\sqrt{\mx + \se}}{\mx} + \dfrac{\sqrt{\fcam + n_\text{pix}^2\se}}{\fcam}}^2\\
    &+ \para{\dfrac{\muncam}{a\times \fcam}}^2\cro{ \dfrac{\Delta \muncam}{\muncam} + \dfrac{\sqrt{\fcam + n_\text{pix}^2\se}}{\fcam}}^2,
\end{aligned}
\end{equation}
where $n_\text{pix}$ is the number of pixels on which the flux is measured and $\Delta \muncam$ the standard-deviation on the noise median $\muncam$ estimation. We get, in nominal conditions, about 1500 photons per pixel and 4.5 electro-photons of read-out noise, the SR estimates is completely limited by the photon noise. At three sigma, we get $\pm$ 1.3 of error on the SR estimation. SR on the reconstructed PSF are directly estimated from the reconstructed OTF, integrated over the spatial frequencies.

The FWHM is determined from fitting five-parameters of the following Moffat function:
\begin{equation} \label{E:moffat}
f(x,y,I_0,\alpha_x,\alpha_y,\beta,\theta)  = I_0 \para{1 + \para{\dfrac{x\cos(\theta) - y\sin(\theta)}{\alpha_x}}^2 + \para{\dfrac{x\sin(\theta) - y\cos(\theta)}{\alpha_y}}^2}^{-\beta},
\end{equation}
where $I_0$ is the PSF amplitude, $\alpha_x$ and $\alpha_y$ coefficients of spreading along $x$ and $y$ directions, $\beta$ the skewness coefficient and $\theta$ the rotation coefficient. The FWHM is then given by:
\begin{equation}
{\text{FWHM}} = \para{2\times (2^{1/\beta}-1)\times (\alpha_x^2 + \alpha_y^2)}^{1/2}.
\end{equation}

We obtained 95~\% of correlation on the SR over 450 data sets using the TS-based method. The reconstruction is affected by a bias of 0.41, which is under the error bars of 1.3 on the H-band images SR. In addition, we get a 1-$\sigma$ standard deviation of 3.52 of SR, over a large swath of observation conditions. In terms of FWHM, we get 92.5\% of correlation, a bias of 2.71 mas and a 1-$\sigma$ standard deviation of 14.9 mas which is 19\% of the diffraction on the WHT. The high correlation obtained on SR and FWHM with the reconstructed PSFs confirm the accurateness of the profiling estimation routines and system calibration. 

From Table~\ref{T:stat}, the RTC-based method achieves very similar results from the TS-based method. On the SR, we get 92.3\% of correlation,a bias of 1.29\% and a 1-$\sigma$ standard-deviation of 4.82\%. On the FWHM, we get 88.3\% of correlation, a bias of 2.94 mas and a 1-$\sigma$ standard deviation of 19.9 mas. As noticed on Fig.~\ref{F:stat}, we get a larger scattering than using the TS-based method.
Bias on the estimates have several sources due to model approximations, as the independence assumption in Eq.~\ref{E:sigepsbertc} and particularly the DM representation using the 36 first Zernike. An important point to notice is the FWHM is similarly well reconstructed with both methods. It means each contributor that spread out the PSF core and contributes significantly to the FWHM, as the static aberrations, are well taken into account.

\subsection{Data set 00h15m36s on September 13th 2013}

We focus now on a particular data set acquired at 00h15m36s on September 13th 2013. \cana was operating in MOAO mode with four LGS and three NGS. During 2048 temporal frames, sampled at 150 Hz, the RTC has provided synchronized WFS telemetry and H-band imaging camera. 

\begin{figure}[H]
\vspace{-4cm}

	\begin{center}
	\includegraphics[scale=.6]{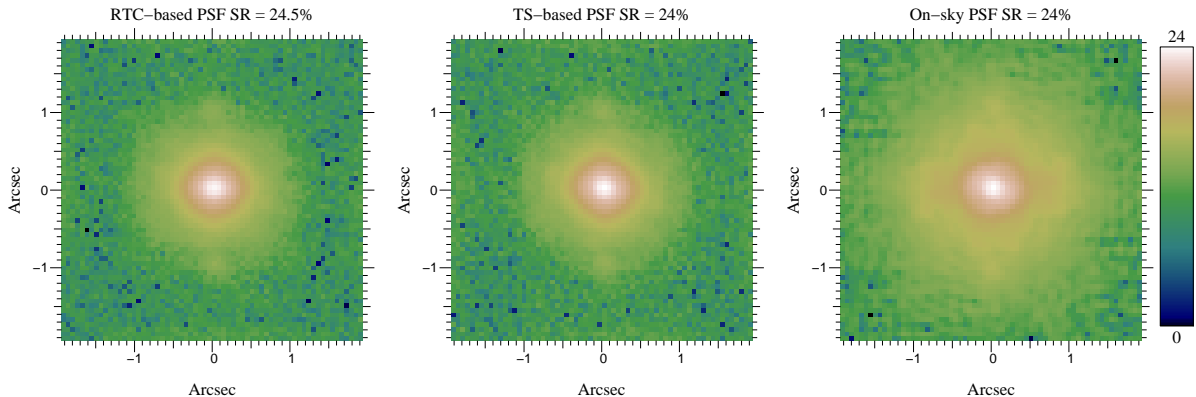}
	\end{center}
	\vspace{-4cm}
	
	\caption{\small{Reconstructed PSF using \textbf{a)} the RTC-based method, \textbf{b)} the TS-based and \textbf{c)} acquired on-sky in H-band~(logarithmic scale). The reconstructed PSFs include an additional Gaussian noise whose the standard-deviation is measured from the corner of the sky PSF, outside the inner tangent circle to the edges of the image.\vspace{.5cm} }}
	\label{F:PSF}
\end{figure}

We report in Fig.~\ref{F:PSF} both sky and reconstructed PSFs. On-sky PSF, we measure a H-band SR of 24.0 and a FWHM of 107.5 mas. With the TS-based method, with a SR of 24.05 and a FWHM of 111.4 mas. Then, with the RTC-based method, we reach 24.5 of SR and 118.8 mas of FWHM. In addition, Fig.~\ref{F:PSF} illustrates the PSF core elongation is well reconstructed whatever the method.

\begin{figure}[h!]
\vspace{-1.5cm}

	\begin{center}
		\includegraphics[scale=.6]{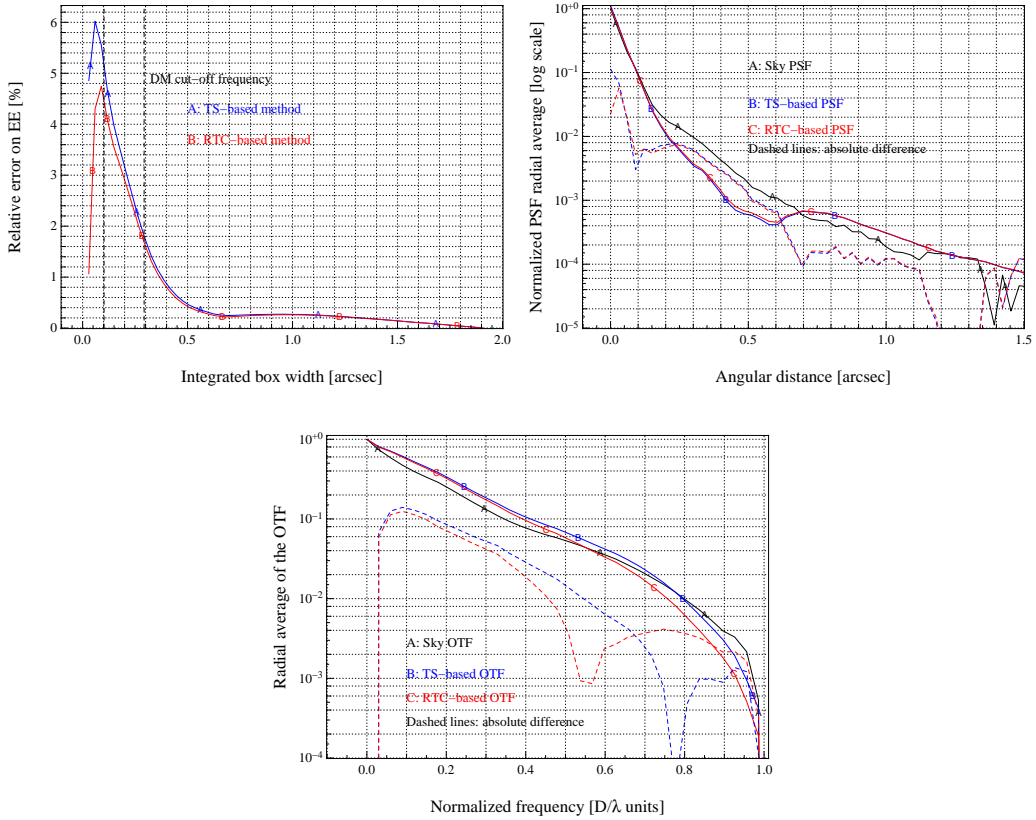}
	\end{center}
	\vspace{-1.5cm}
	
	\caption{\small{\textbf{a):} Relative error on the Ensquared
			Energy computed on the reconstructed PSF, \textbf{b):}
			radial average of both reconstructed/sky PSFs as a function
			of the angular separation from center, \textbf{c): } radial average of both reconstructed/sky OTFs as a function of the normalized frequency.}}
	\label{F:res}
\end{figure}

In Fig.~\ref{F:res} we report the relative error on the Ensquared Energy~(EE), the radial average of the PSFs and the OTFs. One can notice the quality of the reconstruction is not only a matter of SR and FWHM. We stay below  6\% of relative error on the EE. In median value, we get 7.0\% of maximal error on the EE over the 450 processed data for both methods. The signed EE error we get is generally following the curve presented in Fig.~\ref{F:res}: a maximum located in the PSF core and a quick decreasing towards 0. On the OTF, we reconstruct more energy from the low spatial frequencies that translates to a faster decreasing of the reconstructed PSF core than the sky, as we noticed on the radial PSF curves between 0.2 and 0.5 arcsec. 

We believe these mis-matches between reconstructed and sky PSF are first related to assumptions on the DM: the phase is represented by Zernike modes and not by DM influence functions. This effect is minimized when using a reconstruction matrix calibrated on the bench.

\section{Conclusions}
\label{S:conclusions}

We have laid out a PSF-R algorithm tailored to open-loop, tomographic MOAO systems and compared it to a reference method using the Truth-Sensor and to on-sky metrics (SR, FWHM, EE) obtained directly from the science detector.  

We have processed over 450 data sets of telemetry from the \cana MOAO demonstrator. In terms of reconstructed SR, using the RTC-based method, we report a bias and a 1-$\sigma$ standard-deviation of respectively 1.3 and 4.8 of Strehl points, with 92.3\% of correlation between the reconstructed and sky SR. In terms of reconstructed FHWM, we get respectively 2.94 mas and 19.9 mas for the bias and the 1-$\sigma$ standard-deviation, with 88.3\% of correlation using the RTC/TS-based method.  Bias on reconstructed values are lower than errors bars on sky values for both methods on both SR and FWHM. We get also very similar standard-deviation on reconstructed values as well for both methods.

This work shows that fine-detailed system calibration and modeling can be operated to achieve such high levels of correlation on SR and FWHM. We emphasis particularly the reliability and efficiency of tools implemented for turbulence characterization based on the Learn \& Apply approach~(\cite{Martin2016L3S}), and static aberrations calibration, since we are able to reproduce the asymmetric shape of the PSF caused by these static aberrations.

We now plan to improve the reconstruction method by reviewing the approximations and assumptions by evaluating cross-terms using end-to-end simulations and by integrating measurable matrices from the bench on the reconstruction process. We will then extend results to ELT-sized systems.

\acknowledgments 
The research leading to these results received the support of the
A*MIDEX project (no. ANR-11-IDEX-0001-02) funded by the ”Investissements
d'Avenir” French Government program, managed by the French National
Research Agency (ANR). This work is also supported CNRS, INSU, Observatoire de Paris, Universit\'e Paris Diderot-Paris 7 and European Commission (Fp7 Infrastructures 2012-1, OPTICON Grant 312430, WP1). 






\end{document}